\documentclass[structabstract]{aa}  
\usepackage{graphicx}
\usepackage{txfonts}
\usepackage{longtable}
\usepackage{lscape}
\usepackage{natbib}
\def\deltam{\mbox{$\Delta m$}}
\def\deltaz{\mbox{$\Delta z$}}
\def\deltavt{\mbox{$\Delta V_{\rm Ty}$}}

\def\deltaj2{\mbox{$\Delta J_{\rm 2M}$}}
\def\deltah2{\mbox{$\Delta H_{\rm 2M}$}}
\def\deltak2{\mbox{$\Delta K_{\rm 2M}$}}
\def\p/{\mbox{$^1$}}
\def\pp/{\mbox{$^2$}}
\def\ppp/{\mbox{$^3$}}
\def\pppp/{\mbox{$^4$}}
\def\m/{\mbox{$^{-1}$}}
\def\mm/{\mbox{$^{-2}$}}
\def\mmm/{\mbox{$^{-3}$}}
\def\mmmm/{\mbox{$^{-4}$}}
\def\Ms/{\mbox{M$_\odot$}}
\begin{document}
\title{High-resolution imaging of Galactic massive stars with AstraLux}

\subtitle{I. 138 fields with $\delta > -25\degr$}

\author{J. Ma\'{\i}z Apell\'aniz
  \inst{1,2}
  \thanks{Based on data obtained with the 2.2 m telescope at Calar Alto Observatory 
          (CAHA). Also, some images from the NASA/ESA Hubble Space Telescope (HST) were used. 
          The HST data were obtained at the Space Telescope Science Institute, which is operated by the Association 
	  of Universities for Research in Astronomy, Inc., under NASA contract NAS 5-26555.
	  Tables 2 and 3 are available in electronic form at the CDS via anonymous ftp to cdsarc.u-strasbg.fr (130.79.128.5)
          or via http://cdsweb.u-strasbg.fr/cgi-bin/qcat?J/A+A/ .}
         }

   \institute{Instituto de Astrof\'{\i}sica de Andaluc\'{\i}a-CSIC, 
              Glorieta de la Astronom\'{\i}a s/n, 18008 Granada, Spain \\
              \email{jmaiz@iaa.es}
         \and
             Ram\'on y Cajal Fellow.
             }


  \abstract
   {Massive stars have high-multiplicity fractions, and many of them have still undetected components, thus hampering 
   the study of their properties.}
   {I study a sample of massive stars with high angular resolution to better characterize their
   multiplicity.}
   {I observed 138 fields that include at least one massive star with AstraLux, a lucky imaging camera at the 2.2 m
   Calar Alto telescope. I also used observations of 3 of those fields with ACS/HRC on HST to obtain complementary 
   information and to calibrate the AstraLux data. The results were compared with existing information from the Washington 
   Double Star Catalog, Tycho-2, 2MASS, and other literature results.}
   {I discover 16 new optical companions of massive stars, the majority of which are likely to be physically bound to 
   their primaries. I also improve the accuracy for the separation and magnitude difference of many previously known 
   systems. In a few cases the orbital motion is detected when comparing the new data with existing ones and constraints 
   on the orbits are provided.}
   {The analysis indicate that the majority of the AstraLux detections are bound pairs. For a range of separations
   of 0\farcs1-14\arcsec\ and magnitude differences lower than 8, I find that the multiplicity fraction for massive stars
   is close to 50\%. When objects outside those ranges are included, the multiplicity fraction should be considerably
   higher.}


   \maketitle
%

\section{Introduction}

	Massive stars play a crucial role in the dynamical and chemical evolution of galaxies. They are the major source 
of ionizing and UV radiation and, through their huge mass-loss rates, they have a strong mechanical impact on their 
surroundings. 
Massive stars are also important because they are critical contributors to stellar and explosive nucleosynthesis.
The nuclear products are ejected into space in the stellar winds and in the final supernova explosions that put 
an end to the massive stars' lives. Despite their importance, our understanding of these objects and of their 
evolution is still fragmentary due to their relatively small numbers, the existence of unresolved multiple systems, and to 
their concentration along the Galactic plane, where 
extinction affects their detection and the measurement of their distances and other properties.

	A major ongoing project, the Galactic O-Star Spectral Survey, GOSSS 
(PI: J. Ma\'{\i}z Apell\'aniz, see \citealt{Walbetal10a,Gameetal08a}, Sota et al. in prep.), is
currently obtaining ground-based spectroscopy of all known Galactic O stars with $B< 13$ with the purpose of remedying
some of the gaps in our knowledge of Galactic massive stars. All stars are being observed at $R\sim 3000$ to obtain 
accurate spectral classifications, and subsamples are being observed at $R\sim 40\,000$ at multiple epochs to detect
spectroscopic binaries and at $R\sim 1500$ to accurately measure their spectral energy distributions. The detection of
spectroscopic binaries is especially important for massive stars because the multiplicity fraction among them
is especially high \citep{Masoetal98}, a fact that the preliminary GOSSS results are confirming and even increasing in
value. Undetected binaries complicate the study of samples of stars because they introduce biases in the results
and lead to incorrect conclusions. 
The presence of spatially unresolved components can alter the 
observed spectral type and, in some circumstances, yield line-ratio combinations that cannot be present in single O stars
(e.g. different spectral types may be deduced from the He and the N lines, see e.g. \citealt{Walbetal02b}). Therefore, 
correctly characterizing the multiplicity can shed light on objects with such composite spectra. 

	Time-resolved spectroscopy is crucial to detecting short-period multiple systems. However, many massive binaries
are known to have long periods \citep{Masoetal98} and it has been proposed that their period distribution may follow
\"Opik's law \citep{Opik24}, which would lead to essentially all massive stars being born in multiple systems. 
To verify that assertion for large-separation systems, spectroscopy is of little use because the periods involved are 
thousands of years or more and the velocity changes quite small. Therefore, one needs to use high angular-resolution 
methods in order to visually detect companions.
Even so, a gap still exists between the two ranges easily detected by spectroscopic and high angular-resolution methods.
For example, a 30 \Ms/ + 20 \Ms/ in a circular edge-on orbit with $a = 50$ AU at a distance of 1 kpc would have a maximum 
separation of 50 mas and a period of 50 years, both near or beyond the limit of current capabilities. Making the orbit
characteristics, mass ratio, or orientation more unfavorable or placing the system at longer distances will make it even 
harder to detect, to the point that most LMC massive binaries are likely to be currently unknown.

	The likely distribution of binary periods (or separations) among massive stars led me to start a project to 
observe as many massive stars as possible using high-resolution imaging in order to complement the multiple-epoch 
spectroscopy obtained with GOSSS. Such a systematic project is needed to eliminate biases in our knowledge of massive stars. 
Different imaging techniques have been attempted to observe binary systems: speckle interferometry, imaging from space, 
and adaptive optics, among them. Some of the recent attempts are finding previously undetected pairs with large
magnitude differences, \deltam\ \citep{Turnetal08,Maizetal10a}, thus opening a new part of the parameter space. 
In this paper I explore the use of lucky imaging, a technique that, to my knowledge, has not been systematically applied 
to massive stars. In future papers I plan to extend the sample to include several hundred more stars.

\section{Data}

\subsection{AstraLux}

\begin{figure*}
\centering
\includegraphics*[width=\linewidth]{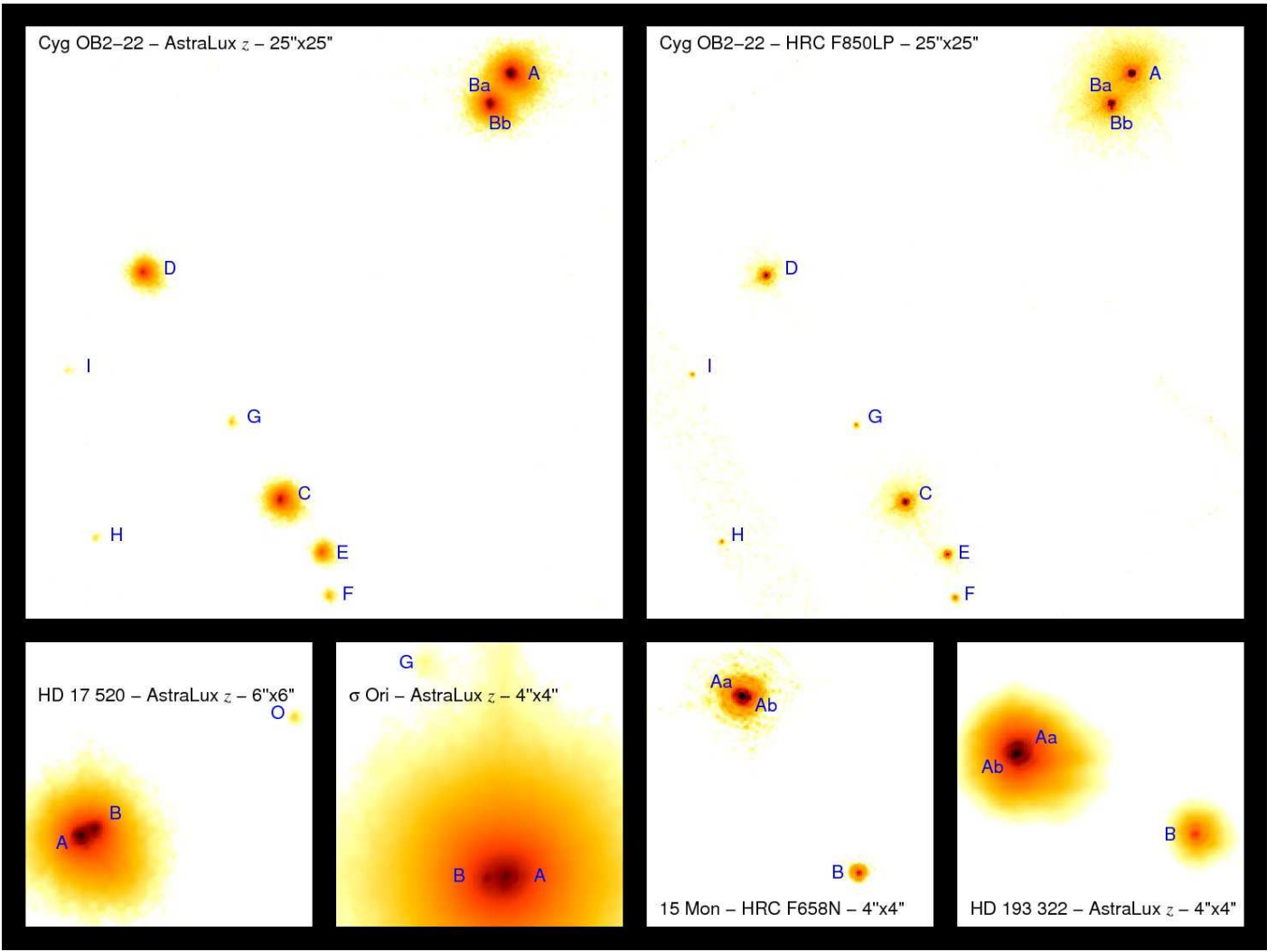}
\caption{Sample AstraLux and HRC data used in this paper. The top two panels show the full AstraLux Cyg OB2-22 field
(left) and its registered HRC equivalent (right). The bottom four panels show sections of four different systems.
In all cases North is up and East is left and a logarithmic stretch is used for the intensity scale. 
See Sota et al. (in prep.) for an enlargement of the top right panel around Cyg OB2-22 A+Ba+Bb.
}
\label{fields}
\end{figure*}

	The main part of the data for this paper was obtained with AstraLux \citep{Hormetal08}, a lucky imaging camera
at the Calar Alto 2.2-m telescope. AstraLux uses an electron-multiplying high-speed CCD capable of obtaining large
numbers (typically, 10\,000) of short-exposure (typically, 30 ms) images of a field of view of 
$24\arcsec\times24\arcsec$. The images are then combined by a pipeline that selects the 1-10\% of the frames with best 
seeing and combines them with a drizzle-type algorithm \citep{FrucHook02}. Under good seeing conditions, AstraLux can 
reach the diffraction limit in the SDSS $i$ and $z$ bands and significantly improve the angular resolution in the rest 
of the visible part of the spectrum. 

	The original goal of this project was to observe with AstraLux the 185 O stars with $\delta > -25\degr$ in the
Galactic O-Star Catalog of \citet{Maizetal04b}. 
To that purpose, I observed for ten nights between 12 November 2007 and
10 May 2009, in all cases using 30 ms exposures in the SDSS $z$ band and adjusting the detector amplifier to maximize 
the S/N without reaching saturation. Unfortunately, I could observe for less than half of the allocated time because of
bad weather and poor seeing, leaving only 138 observed fields. Those include a few
massive stars not in the original catalog but present in its second version \citep{Sotaetal08}, as well as some WR stars. 
In most cases, a single massive star was known to exist in the field and the AstraLux field was centered on it.
For some fields, the primary object was positioned off-center in order to include previously known companions. The
effective search radius for companions as measured from the brightest star in a given field is 13\farcs6, but given the 
uneven coverage in position angle some companions were found out to 26\arcsec. 
The effective central wavelength for the observations is 9115 \AA\ (with variations of $\pm 5$ \AA\ as a function of 
the airmass) and the bandpass is 1250 \AA\ (defined as the wavelength range containing 68\% of the total throughput).

	Each field was processed at Calar Alto using the existing instrument pipeline. In order to maximize angular
resolution, I selected for the output images just the 1\% of the frames with better seeing. Observations of the
Orion Trapezium ($\theta^1$ Ori) were used to measure the geometric distortion of the images, with the reference
positions of the Trapezium stars measured from an archival WFPC2/HST image. Given the relatively small number of stars
visible in the AstraLux Trapezium images, only the linear terms of the geometric distortion were obtained. The fit
residuals for the Trapezium stars and AstraLux observations of a comparison field, Cyg OB2-22, for which I had ACS/HRC
images (see below), were used to estimate the uncertainties in separation and position angle of the observations. The 
geometric distortion of the pipeline output images was then removed to produce the final astrometrically corrected images
(Fig.~\ref{fields}).

	The AstraLux image of a single star observed under good ($\sim$0\farcs7) or intermediate ($\sim$1\farcs0) seeing 
conditions consists of a core that is near-diffraction limited and a halo that is determined by the seeing conditions.
The PSF does not appear to show significant variations over the field of view.
Since the goal here is to accurately determine the separations and magnitude differences (\deltam, in this case, \deltaz)
for the pairs observed in each field, I wrote a PSF-fitting program in IDL to simultaneously calculate the positions and
fluxes of all detected stars. The PSF itself was fitted using a model that consists of a Gaussian core and a Moffat
halo \citep{Moff69}. For both components (core and halo), I allowed for the profile to have an elliptical shape in the image plane.
When determining the fluxes for each component, I considered both the random uncertainties produced by the PSF-fitting
program and the systematic uncertainties resulting from the differences between the real PSF and the fitted profile. The
reported uncertainties in \deltam\ include both effects. Also, for those cases consisting of a close small-\deltam\ 
binary and a third bright component I did aperture photometry checks to evaluate possible systematic effects in the
PSF fitting. 

\begin{table}
\caption{Systems with a single component in the AstraLux data.}
\label{single_data}
\centering
\begin{tabular}{llll}
\hline\hline
\multicolumn{1}{c}{System name} & \multicolumn{1}{c}{System name} & \multicolumn{1}{c}{System name} & \multicolumn{1}{c}{System name} \\
\multicolumn{4}{c}{} \\
\hline
\multicolumn{4}{c}{} \\
   HD 225\,160 &     HD 18\,409 &     HD 46\,573 &         MY Ser \\
    BD +60 261 &         CC Cas &     HD 46\,966 &    HD 186\,980 \\
    HD 12\,323 &          X Per &       V689 Mon &    HD 188\,209 \\
    HD 12\,993 &      $\xi$ Per &     HD 47\,417 &          9 Sge \\
      V354 Per &   HDE 237\,211 &     HD 48\,099 &    HD 192\,639 \\
    HD 14\,434 &   $\alpha$ Cam &     HD 52\,266 &    HD 195\,592 \\
    HD 14\,442 &         AE Aur &     HD 53\,662 &     Cyg OB2-11 \\
    HD 14\,633 &   HDE 242\,908 &     HD 53\,975 &    HD 199\,579 \\
    HD 14\,947 &    BD +34 1058 &     HD 54\,662 &         68 Cyg \\
    HD 15\,137 &    BD +39 1328 &     HD 54\,879 &    HD 207\,198 \\
    BD +60 497 &     HD 37\,737 &     HD 55\,879 &         LZ Cep \\
    HD 15\,570 &      V1382 Ori &     HD 57\,236 &         19 Cep \\
    HD 15\,642 &     HD 41\,997 &     HD 57\,682 &  $\lambda$ Cep \\
    HD 15\,629 &     HD 42\,088 &     HD 60\,848 &         10 Lac \\
    BD +60 513 &     HD 45\,314 &     HD 93\,521 &    HD 216\,532 \\
    BD +62 424 &     HD 46\,149 &    $\zeta$ Oph &    HD 216\,898 \\
    HD 16\,691 &     HD 46\,223 &    Herschel 36 &    HD 218\,915 \\
    HD 17\,603 &     HD 46\,485 &          9 Sgr &    BD +60 2522 \\
\multicolumn{4}{c}{} \\
\hline
\multicolumn{4}{c}{} \\
\end{tabular}
\end{table}

	Of the 138 fields observed, I detected a single star in the AstraLux images in 72 fields and two or more stars 
in the other 66. Given the PSF complexity, I found that visual inspection was the most effective detection mechanism.
The systems with a single detected star are given in Table~\ref{single_data}. For the systems where 
two or more stars are detected in the AstraLux data, the measurements for the detected pairs are shown in 
Table~\ref{pair_data}. In all cases I searched the Washington Double Star Catalog (WDS, \citealt{Masoetal01}, the used 
version was downloaded on 10 December 2009) and the Tycho-2 \citep{Hogetal00a} and 2MASS \citep{Skruetal06} databases for 
possible known counterparts of the detected sources.

	Table~\ref{pair_data} lists a total of 142 pairs, of which 33 were not listed in the WDS. For the new components I have
followed a nomenclature consistent with that in the WDS. Seventeen of the 33 pairs not in the WDS had both components in 
the 2MASS database, leaving 16 components that will be considered strictly new in this paper.

\subsection{ACS/HRC}

	Three of the fields, Cyg OB2-22, Cyg OB2-7, and 15 Mon, observed with AstraLux were also observed with ACS/HRC 
on HST\footnote{Note that there are relatively few HST images of Galactic bright stars with an exposure time short
enough not to saturate the central part of the object PSF.}.
The Cyg OB2 observations are part of GO program 10602, of which the author is the P.I., and that had as its goal
the search for companions around the earliest Galactic O stars \citep{Maizetal07}. The 15 Mon observations were obtained 
as part of an HST calibration program.

	HRC was a 1024$\times$1024 CCD 
sensitive in the 2000-10\,000 \AA\ range with an average pixel scale of 0.027\arcsec/px, hence providing a similar field
of view and angular resolution as AstraLux at the same wavelengths where the latter works best. The operative advantages 
of HRC over AstraLux are its better Strehl ratio, its lack of dependence on atmospheric conditions, and its better 
contrast because of its single-frame operation. A comparison between the two instruments can be established from the top
two panels in Fig.~\ref{fields}. Note how the HRC F850LP halo is significantly less intense than that of the AstraLux
$z$ image (both filters are essentially equivalent). The HRC halo in the F850LP images is not an optical effect but is 
instead caused by scattering in the detector \citep{acs}. At shorter wavelengths the effect disappears, as it can be see
in the extremely compact PSF of 15 Mon in one of the lower panels of Fig.~\ref{fields}.

	The HRC individual images were processed through the {\tt calacs} pipeline and combined with {\tt multidrizzle} 
to generate geometrically-corrected mosaics. A crowded-field photometry package, JMAPHOT, written by the author for HST 
data, was used to [a] search for sources in the multidrizzled mosaics, and [b] extract the photometry from the individual
exposures. The observed photometry was corrected for charge transfer inefficiency effects \citep{acs}.
The same sources found in the AstraLux images were detected in the HRC data plus another object (Cyg OB2-7
E) that fell just outside the AstraLux field and just inside the HRC field. JMAPHOT uses the PSFs and geometric
distortions of \citet{AndeKing04} for the HRC, which provide excellent photometric and astrometric accuracies. The 
results for the three HRC fields are shown in Table~\ref{hrc_data}. 

	Note that for the two Cyg OB2 fields only the F850LP data will be used in this paper. We plan to use the 
information in the rest of the observed filters in a future work.

\begin{table*}
\caption{ACS/HRC measurements of systems with multiple components.}
\label{hrc_data}
\centering
\begin{tabular}{lcccr@{$\pm$}lr@{$\pm$}lr@{$\pm$}l}
\hline\hline
\multicolumn{1}{c}{System name} & \multicolumn{1}{c}{Filter} & \multicolumn{1}{c}{Pair} & MJD & \multicolumn{2}{c}{Separation} & \multicolumn{2}{c}{Orientation} & \multicolumn{2}{c}{$\Delta m$} \\
                                &                            &                          &     & \multicolumn{2}{c}{(\arcsec)}  & \multicolumn{2}{c}{(degrees)}   & \multicolumn{2}{c}{(mag.)}     \\
\multicolumn{10}{c}{} \\
\hline
\multicolumn{10}{c}{} \\
15 Mon     & F658N  & Aa-Ab & 53669.96 & $ 0.089$ & $0.001$ & $247.90$ & $0.57$ & $1.62$ & $0.01$ \\
           &        & Aa-B  & 53669.96 & $ 2.969$ & $0.001$ & $213.29$ & $0.01$ & $3.08$ & $0.01$ \\
Cyg OB2-22 & F850LP & A-Ba  & 53733.14 & $ 1.521$ & $0.001$ & $146.19$ & $0.03$ & $0.59$ & $0.01$ \\
           &        & Ba-Bb & 53733.14 & $ 0.216$ & $0.001$ & $181.48$ & $0.25$ & $2.34$ & $0.01$ \\
           &        & A-C   & 53733.14 & $20.404$ & $0.001$ & $152.37$ & $0.01$ & $1.15$ & $0.01$ \\
           &        & C-D   & 53733.14 & $11.191$ & $0.001$ & $ 31.37$ & $0.01$ & $0.57$ & $0.01$ \\
           &        & C-E   & 53733.14 & $ 2.821$ & $0.001$ & $218.41$ & $0.02$ & $1.73$ & $0.01$ \\
           &        & E-F   & 53733.14 & $ 1.850$ & $0.001$ & $189.82$ & $0.06$ & $1.89$ & $0.04$ \\
           &        & C-G   & 53733.14 & $ 3.853$ & $0.001$ & $ 32.41$ & $0.02$ & $4.02$ & $0.02$ \\
           &        & C-H   & 53733.14 & $ 7.864$ & $0.001$ & $102.32$ & $0.01$ & $4.36$ & $0.06$ \\
           &        & C-I   & 53733.14 & $10.404$ & $0.001$ & $ 58.88$ & $0.01$ & $4.56$ & $0.05$ \\
 Cyg OB2-7 & F850LP & A-B   & 53736.28 & $18.604$ & $0.001$ & $104.70$ & $0.01$ & $2.05$ & $0.01$ \\
           &        & A-C   & 53736.28 & $15.682$ & $0.001$ & $106.50$ & $0.01$ & $5.80$ & $0.02$ \\
           &        & A-D   & 53736.28 & $20.399$ & $0.001$ & $ 79.59$ & $0.01$ & $5.54$ & $0.01$ \\
           &        & A-E   & 53736.28 & $24.987$ & $0.004$ & $104.18$ & $0.01$ & $8.19$ & $0.15$ \\
\multicolumn{10}{c}{} \\
\hline
\multicolumn{10}{c}{} \\
\end{tabular}
\end{table*}

\section{Individual systems}

	In this section I analyze the systems where there is new significant information compared to the literature. 
This could be because [a] new components are detected, [b] previously detected components are not seen, [c] the 
photometry or the astrometry are discrepant with the literature values, or [d] a relative proper motion is detected.

\subsection{HD 5005}

	The measured \deltaz\ values appear to be inconsistent with the Tycho-2 and 2MASS results (all four components
are O stars with similar extinctions, see Sota et al. in prep., so they should have similar colors). For Tycho-2, a
possible explanation is that $\approx 0.3$ mag from A has been incorrectly assigned to B, hence decreasing the 
\deltam\ between A and the other three components. In the case of 2MASS, B is not detected and its flux appears to
have been distributed between A and C. The results of Sota et al. (in prep.) for the flux ratios between components in the 
$B$ band from long-slit spectroscopy agree with the \deltaz\ values within less than 0.2 mag, again indicating a
problem with the Tycho-2 photometry.

	The 2MASS and Tycho-2 astrometry for HD 5005 appear to be relatively poor, with offsets of
several tenths of an arcsecond.

\subsection{HD 8768}

	The $H$ and $J$ 2MASS photometry for the C component are only upper detection limits, probably because of
contamination from the unresolved A-B pair. Also, the \deltak2\ for A-C is likely overestimated because the flux from B
is included in A.

\subsection{BD +60 499}

	The 2MASS position for B is offset by several arcseconds (as can be seen by overplotting the point source
catalog on top of e.g. the $J$ image). Also, some of the flux from A is apparently assigned to B, leading to erroneous
results for the 2MASS \deltam\ values.

\subsection{HD 16\,429}

	The Tycho-2 and 2MASS \deltam\ values for Aa-B likely refer to (Aa+Ab)-B but the effect should be small, 
given the large \deltaz\ for Aa-Ab. The \deltaz\ uncertainties are especially large because of a poor PSF fit.

\subsection{HD 16\,832}

	A new component (C) is detected. 

\subsection{HD 17\,505}

	The 2MASS \deltam\ values likely refer to (A+B) rather than to A alone.

\subsection{HD 17\,520}

A new component (O) is detected (Fig.~\ref{fields}). The 2MASS \deltam\ values for A-C likely refer to (A+B)-C.

\subsection{BD +60 586}

	A new component (F) is detected.

\subsection{HD 18\,326}

	A new component (F) is detected. The 2MASS \deltam\ values likely refer to (A+B) rather than to A alone.

\subsection{NSV 1458 - SZ Cam}

	This system (WDS 04078+6220) has generated a lot of confusion in the past regarding the use of HD numbers to 
identify each one of the two stars (\citealt{Loreetal98}, Sota et al. in prep.). I observed it with AstraLux using two fields: one 
that included NSV~1458 (the A component in WDS) plus B, C, and D; and another with the two components of SZ Cam (Ea and
Eb in WDS) plus A, B, and C. 

	The Ea-Eb pair is one of the three in this paper that I resolve with a separation of less than 0\farcs15. The 
uncertainty in \deltaz\ is much smaller for the Ea-Eb pair than for the other pairs involving Ea because it depends only 
on the fitting of the stellar cores. For the other pairs (e.g. Ea-A) there is a substantial residual in the fitting due 
to the assignment of the halo contribution to either Ea or Eb. On the other hand, it is possible to do simple aperture 
photometry and obtain a much more reliable value of \deltaz\ for (Ea+Eb)-A of $-0.28\pm0.01$ mag. 
	
	Ea is an eclipsing spectroscopic binary \citep{Loreetal98} and I must have caught the system in an eclipse, 
since the \deltaz\ value for (Ea+Eb)-A is too negative when compared with the Tycho-2 and 2MASS photometry, which
do not resolve Ea-Eb. 
Indeed, the ephemerides of \citet{Loreetal98} and \citet{Gordetal07} place the observation at a 
phase distance of just 0.02-0.03 of the secondary eclipse, which is almost as deep as the primary one. 
The depth of the
secondary eclipses given by \citet{Loreetal98} are $\sim 0.2$ mag, which is in the range
needed to make the result here consistent with the Tycho-2 and 2MASS values. Another point in favor of the system being 
in an eclipse is that the AstraLux photometry gives Ea and Eb as having very similar magnitudes, while the WDS
gives a \deltam\ for the pair of 0.50 mag. New observations with AstraLux at different epochs should be able to produce 
a light curve.

	At first sight, the astrometry I measure for Ea-Eb indicates a separation somewhat greater than the one predicted 
by the \citet{Gordetal07} orbit. However, since there is a significant residual in the PSF fitting and the error bars 
here are relatively large, it is not clear whether the two results are inconsistent or not.

\subsection{AE Aur (= HD 34\,078)}

	The Ab component is not detected because the measured separation of 0\farcs35 and the expected \deltaz\ derived 
from \citet{Turnetal08} are likely to be outside the parameter space where AstraLux can identify individual components
unless seeing is exceptionally good (see Fig.~\ref{sep-dm}).

\subsection{HDE 242\,926}

	A new component (B) is detected.

\subsection{HDE 242\,935}

	The 2MASS \deltam\ values for A-C and A-D likely refer to (A+B)-C and (A+B)-D, respectively.

\subsection{$\theta^1$ Ori (= HD 37\,020 + 1 + 2 + 3)}

	The 2MASS photometry for the Trapezium stars is inconsistent with the AstraLux and Tycho-2 values and 
with the NIR speckle holography results of \citet{Petretal98}, possibly because of the combined effect of nebular 
contamination and crowding. I was unable to separate A or C into their respective components. 

	In the case of
B, I easily resolved Ba and Bb but could not detect Bc or Bd \citep{Scheetal03}. The large magnitude difference
in Ba-Bb confirms the red colors of Bb \citep{Scheetal03}. 
Note that the measured \deltaz\ for C-Ba is greater
than expected: Ba is a known eclipsing binary \citep{Steletal05} and I must have observed it during an eclipse. 
Indeed, the ephemerides of \citet{Vitr08} provide a phase distance of less than 0.01 between the nearest primary 
eclipse and the time of observation. Although the O--C value could be non-negligible,
the eclipse is rather long and its profile near minimum quite flat \citep{Wolf94}, so the AstraLux 
observation must have experienced a large magnitude difference with respect to the uneclipsed case. The curves of 
\citet{Wolf94} yield a magnitude difference of $\sim 0.5$, which is consistent with the data here.

	A new component (Z) not present in either the WDS, 
Tycho-2, or 2MASS catalogs is detected; note, however, that it had already been identified as TCC 63 by 
\citet{McCaStau94}.

\subsection{$\sigma$ Ori (= HD 37\,468)}

	The AstraLux photometric uncertainties are relatively large because of the combination of a poor PSF (see 
Fig.~\ref{fields}, which shows a PSF elongation along a position angle of $\approx$160\degr) and the 
proximity of the A and B components. The Tycho-2 and 2MASS \deltam\ values likely refer to (A+B) rather than to A 
alone and they may include G as well, also known as IRS1 (see \citealt{Caba07}, \citealt{Bouyetal09}, and 
Fig.~\ref{fields}). The separation and orientation for A-B are consistent with the ephemerides of \citet{Turnetal08}.

\subsection{HD 37\,366}

	The 2MASS \deltam\ values for A-C likely refer to (A+B)-C.

\subsection{HD 46\,202}

	The 2MASS \deltam\ are apparently incompatible with the AstraLux results (unless E is very red), possibly due 
to crowding. Note that WDS 06321+0458 A is HD 46\,180, located 49\arcsec\ towards the NW and not included in the
AstraLux field of view.

\subsection{HD 46\,223}

	The B component of \citet{Turnetal08} was not detected (see Fig.~\ref{sep-dm}), likely because of a non-optimal 
seeing of 1\farcs04.

\subsection{V689 Mon (= HD 47\,432)}  

	The B component of \citet{Turnetal08} was not detected (see Fig.~\ref{sep-dm}), likely because of a seeing of 1\farcs08.

\subsection{HD 47\,032}

	A new component (B) is detected.

\subsection{HD 47\,129}

	The C component of \citet{Turnetal08} was not detected (see Fig.~\ref{sep-dm}), likely because of a seeing of 1\farcs04.

\subsection{15 Mon (= HD 47\,839)}

	Aa-Ab is another one of the closest pairs analyzed in this paper. The \deltavt\ value for Aa-B likely
refers to (Aa+Ab)-B. 

	The Aa-Ab orbit has been followed for some time now \citep{Giesetal97,Masoetal09} and the most recently
published orbit \citep{Cvetetal10} is significantly wider than the previous one of \citet{Giesetal97}. In this paper I 
publish two new points, one for October 2005 (HRC, Fig.~\ref{fields}) and another one for January 2008 (AstraLux). The 
2005 value is consistent with the \citet{Cvetetal10} orbit but the 2008 value is not, indicating that Ab is moving faster and 
in a wider orbit. If one looks at Fig.~1 in \citet{Cvetetal10} it is clear that their last two measurements are also 
discrepant. The AstraLux results point in the direction of the outer one of those two measurements being closer to the 
truth than the inner one. 

	The astrometry for Aa-B and the photometry for the two pairs show a good agreement between HRC and AstraLux
(within the expected color term between F658N and $z$), hence providing a confirmation of the AstraLux calibration.

\subsection{HD 52\,533}

	The 2MASS \deltam\ values for A-E and A-F likely refer to (A+B)-E and (A+B)-F, respectively.

\subsection{HD 193\,322}

	Aa-Ab is the closest pair that I am able to identify in the sample in this paper, with a separation of only 
55$\pm$16 mas. The identification is possible thanks to the existence of the bright B companion, which allows for an
accurate PSF modeling (see Fig.~\ref{fields}, where B appears to have good circular symmetry and Aa+Ab appear as an
elongated single object). The separation and orientation are compatible with the \citet{Masoetal09} results. The
\deltaz, which is rather uncertain, is quite different to the \citet{Masoetal09} value for \deltam\ but 
within 2 sigma of the \cite{McKietal98} result. Note that the \deltavt\ value refers to (Aa+Ab)-B.

\subsection{Cyg OB2-5 (= V279 Cyg = BD +40 4220)}

	A new component (D) is detected. The 2MASS \deltam\ values for A-C likely refer to (A+B)-C. 
Note that the component referred to as D in \cite{Kennetal10} is B here and in the WDS.

\subsection{Cyg OB2-22}

	There are five components detected (Bb, E, F, G, and I) in the AstraLux data that are not present in the WDS
or 2MASS (this system is too dim for Tycho-2), for a total of 10 components (Fig.~\ref{fields}). 
Of the five new components, the last four are detected by \citet{MassThom91} as objects 420, 807, 814, and 426, respectively. 
The 2MASS \deltam\ values likely refer to (A+Ba+Bb) rather than to A alone. The same ten components are detected with ACS/HRC. 

	The AstraLux \deltaz\ values are in general consistent with the HRC F850LP results with two caveats. [1] The 
AstraLux value for Ba-Bb has a very large uncertainty because Bb is close to the detection limit for its \deltaz\ and 
separation. On the other hand, HRC cleanly separates the two. [2] C appears to be too dim in the AstraLux photometry 
with respect to HRC by 0.05-0.10 magnitudes. One possibility is that C is variable. 

	With the exception of Ba-Bb, the AstraLux separations and orientations for all pairs are consistent with the HRC 
results, thus validating the Trapezium-based AstraLux astrometric calibration used in this paper (no significant proper
motions are expected in the $\approx$2 years that lapsed between the HRC and AstraLux observations). The Ba-Bb 
separation may have increased while maintaining a near-constant orientation, but such a motion is only a 2$\sigma$
detection in a close system with a relatively large \deltaz, so it may not be real. 

\subsection{Cyg OB2-7}

	A new component (C) is detected in both the AstraLux and the HRC data and another one (E) in the HRC images 
(E is just outside the AstraLux field of view). The astrometry for AB, AC, and AD show a good agreement between 
HRC and AstraLux, hence providing another confirmation of the AstraLux calibration. The photometry also shows a good
agreement for AB and AC but not for AD. A possible explanation is that D is variable. 

\subsection{HD 206\,267}

	The Tycho-2 \deltam\ values appear to be incompatible with the AstraLux \deltaz, the 2MASS \deltam, and 
the published $BV$ photometry \citep{Nico78}. The B component of \citet{Turnetal08} is undetected because of the very bad 
seeing (1\farcs85, see Fig.~\ref{sep-dm}).

\subsection{DH Cep (= HD 215\,835)}

	A new component (B) is detected.

\subsection{HD 218\,195}

	A new component (C) is detected. The 2MASS \deltam\ values for A-D likely refer to (A+B+C)-D.

\section{Discussion}

\subsection{Overall strategy}

	This paper is the first step of a long-term project to study the multiplicity of massive stars. The project will
include further high-angular resolution imaging, multiple-epoch $R\sim 40\,000$ optical spectroscopy \citep{Gameetal08a},
a recalibration of the intrinsic colors for hot stars and the extinction law \citep{MaizSota08}, and a combination of
spectroscopic and trigonometric parallaxes \citep{Maizetal08c}. Regarding high-resolution imaging, the goal is to
eventually observe $\sim 500$ massive stars with AstraLux and, if a similar instrument becomes available in the Southern
hemisphere, to observe another $\sim 500$ stars from there, in all cases using the Galactic O-Star Catalog 
\citep{Maizetal04b,Sotaetal08} to select the sample. 

\begin{figure}
\centering
\includegraphics*[width=\linewidth]{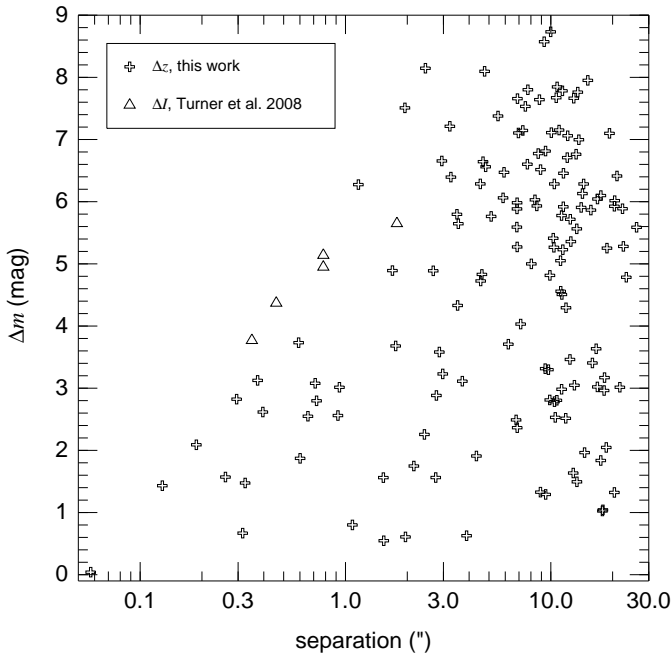}
\caption{Separation-\deltam\ plot for the pairs detected with AstraLux, where pairs are selected with one of the
components being always the brightest star in the field. As a reference, the five pairs with $\deltam < 9.0$ detected by 
\citet{Turnetal08} but not seen in the AstraLux data are also plotted.}
\label{sep-dm}
\end{figure}

	Literature results will also be included in our final analysis. In that respect, it is
useful to compare the sample size and AstraLux capabilities with previous large-scale surveys. 
\citet{Masoetal98,Masoetal09} observed $\sim 400$ O and WR stars with speckle interferometry, which has the advantage of
being able to reach down to lower separations (0\farcs03) but the disadvantages of lower maximum separations (1\farcs5)
and dynamic range ($\deltam \sim 3$). \citet{Nelaetal04} used the Fine Guidance Sensor (FGS) 1r aboard HST as an 
interferometer to observe 23 OB stars in the Carina Nebula. FGS can reach to even smaller separations than speckle
interferometry but reaches out only to $\sim 0\farcs5$. \citet{Turnetal08} observed a sample similar in size to the 
one here with $I$-band adaptive optics. Their instrument explores a similar region of the separation-\deltam\ plane as 
AstraLux, with a slight advantage in \deltam\ ($\sim 1.5$ mag at separations of 3-6\arcsec, see Fig.~\ref{sep-dm} where 
18 objects with $\deltam \sim 10$ detected by \citealt{Turnetal08} lie outside the plotted range) and disadvantages in 
separation range (6\arcsec\ instead of 12--18\arcsec) and \deltam\ precision. 

\subsection{Completeness}

	Fig.~\ref{sep-dm} shows the separation and \deltaz\ of all of our detected pairs, where values are measured
from the brightest star, not for the pairs listed in Table~\ref{pair_data} (the same criterion applies to subsequent 
figures). There is an obvious detection
limit that runs from $\approx 0\farcs1$ for small \deltaz\ to $\approx 2\arcsec$ for 
\deltaz\ = 8 magnitudes\footnote{Note that the point near the lower left corner, which corresponds to HD 193\,322 Aa-Ab,
is a special case that is detected because the existence of nearby, bright B allows for a better-than-normal PSF modeling,
see above.}. The location of
the detection limit is not the same for all of the observations because it is seeing-dependent. A consequence of this
dependence is that one of the five AstraLux non-detections on Fig.~\ref{sep-dm} (HD 206\,267) is surrounded by AstraLux 
detections due the very bad seeing of that particular AstraLux observation. The AstraLux detections also appear to 
become affected by incompleteness at large distances for $\deltaz > 8$. From Fig.~\ref{sep-dm} it can be estimated that
if the \deltam\ distribution is independent of separation, then $\sim 30$ stars with separations between 0\farcs1 and
2\arcsec\ and $\deltaz < 8$ are undetected by AstraLux. 

\begin{figure}
\centering
\includegraphics*[width=\linewidth]{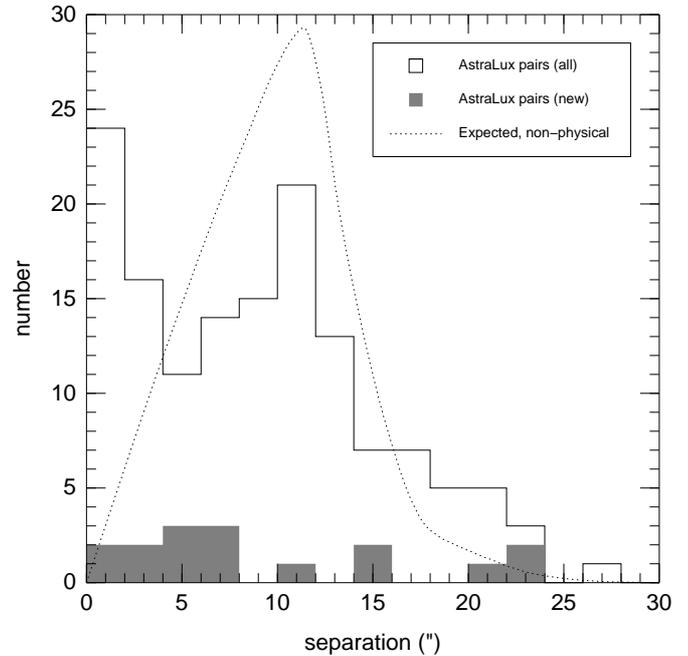}
\caption{Separation histogram for all and for new pairs detected with AstraLux. The dotted line indicates the expected
shape for a population of non-physical pairs uniformly distributed in the plane of the sky. 
As in the previous figure,
pairs are selected with one of the components being always the brightest star in the field. The new pairs with large
separations belong to systems with many components; if measured from the nearest companion, their separations would be 
considerably smaller.
}
\label{sep}
\end{figure}

\subsection{Bound or not?}

	A basic question that needs to be answered is how many of the observed pairs are [a] real, bound pairs and how 
many are either [b] stars located in the same cluster but not bound to the central massive stars or [c] chance
alignments from stars located along the Galactic line of sight (foreground or background). To solve this issue with
certainty, one would need to measure the proper motions and velocities of each component, a task that is beyond the scope
of this first paper (but that will be eventually tackled within the GOSSS project). Here I attempt a statistical
approach with the help of Fig.~\ref{sep}, which shows the separation histogram for the AstraLux-detected pairs. It also
shows the expected non-normalized distribution for pairs built by associating the central massive star with a population 
of stars uniformly distributed over the field of view. Such an expected distribution initially increases linearly with
separation and then turns around where the detector edge is reached. It was built taking into account the distribution
of the central stars on the AstraLux CCD\footnote{As previously mentioned, for some targets the brightest star was
located off-center in order to observe previously-known components. For some cases with a large separation
the pair was aligned close to the NW-SE or NE-SW directions.}.

	A comparison between the observed separation of the AstraLux pairs and the expected distribution described above 
shows an important difference: a strong peak is present at small separations. Furthermore, as previously discussed, the 
first bin that defines this peak (0\arcsec-2\arcsec, in practical terms the lower limit is something like 0\farcs1) should 
be strongly affected by incompleteness so its real height should be roughly double what is shown in Fig~\ref{sep}. 
The conclusion is that this peak is formed by real, bound pairs and not by cluster or Galactic chance alignments.

	On the other hand, for separations beyond 6\arcsec\ the observed distribution is relatively well-approximated 
by the expected distribution formed by a uniform population. There may be an excess of objects with large
separations ($> 16\arcsec$) but this can be simply a selection effect due to the specific positioning of the targets
when the existence of a component was known in advance. Note, however, that the similarity of the two distributions does
not allow us to distinguish between options [b] and [c] above or even a specific case of [a], a population of bound
pairs with large separations spread over a broad range. In order to differentiate between the three we can take
advantage of the fact that for separations beyond 8\arcsec\ one would expect 2MASS detections to be relatively complete 
(i.e. unaffected by the presence of a bright nearby star) for $\deltam < 9$. This can be seen from the fact that in the
separation range 8\arcsec-14\arcsec\ forty-six\footnote{In thirty-three different fields, some with two or three pairs.}
out of forty-nine the AstraLux detections are also in 2MASS (two more are in the WDS and the last one is a new 
detection) and can also be deduced from Fig.~6 in \citet{Turnetal08}. Therefore, we can use 2MASS to estimate the 
fractions of the pairs in the peak around 11\arcsec\ of separation that are either bound systems, 
cluster members, or unrelated Galactic alignments.

	I start by defining three separation ranges for 2MASS: [1] the stellar system (8\arcsec-14\arcsec), [2] the cluster
(1\arcmin-4\arcmin), and [3] the Galactic fore/background (30\arcmin-40\arcmin)\footnote{The limits for [2] and [3] are
realistic but somewhat arbitrary and the method here assumes that the surface density for cluster and Galactic members does
not vary between the center and the maximum separation in each case, making this estimate a rather unsophisticated one. 
Nevertheless, the results below were found to be robust with respect to small changes in the limits.}. 
As previously mentioned, AstraLux detects 46 objects (in 33 fields) in the stellar-sytem range.
To that number I apply a geometric correction that accounts for the incomplete angle coverage to determine 
that AstraLux would detect $\sim 61$ sources between 8\arcsec\ and 14\arcsec\ if it had covered that separation range in full. 
On the other hand, there are 106 2MASS detections (in at least one band) in that range for the same fields, indicating the 
existence of a detection limit in magnitude for Astralux with respect to the 2MASS data (i.e. 2MASS is deeper for those
separations).  I then [a] calculate the fraction of AstraLux-detected stars for each of the three 2MASS bands ($J$, $H$, 
and $K_{\rm s}$); [b] convert the result to 2MASS magnitude limits for the AstraLux detections; and [c] obtain the 2MASS 
stellar density above those magnitude limits at the other two separation ranges (cluster and Galactic) centered on each 
of the 33 AstraLux fields. From those values and assuming a Poisson distribution, I 
obtain the probability that the observed AstraLux pair or pairs in the stellar system range belongs to either [a], [b],
or [c]\footnote{The probability for [a] and [c] is always found to be non-zero. This is not always true for [b], since
some O stars do not belong to a cluster.}. I then sum over the 33 AstraLux fields and I obtain the percentage 
of pairs that falls in each category. Since the process is repeated for the three 2MASS bands, I derive a mean value and
a dispersion for each of the percentages.

	The result of the above procedure is that for separations of 8\arcsec-14\arcsec, $78\pm 6\%$ of the pairs are 
bound, $10\pm 2\%$ are chance alignments with a cluster member, and 
$12\pm 4\%$ are chance alignments with a non-cluster (Galactic
fore/background) member. These numbers indicate a rather large proportion of bound pairs in the AstraLux sample. The
fraction of bound stars is slightly greater than the values reported by \citet{Turnetal08} but it also represents a
different range in separation and \deltam\ (more specifically, their sample includes many stars with $\deltam\sim 10$
mag). It is worth mentioning that there are no dense clusters with O stars above a declination of $-25\degr$ (NGC 3603, 
Trumpler 14, and NGC 6231 are all further south, for example); in those cases one would expect a higher fraction for
chance alignments with cluster members \citep{Maiz08a}. Also, it should be remembered that these values are preliminary 
pending a larger sample and proper motion and radial velocity studies. 

\subsection{The multiplicity fraction}

	The quest for the multiplicity fraction is the ultimate goal of a work like this. Unfortunately, at this stage
there are many limitations for the numbers that can be provided based on the current data.

\begin{figure}
\centering
\includegraphics*[width=\linewidth]{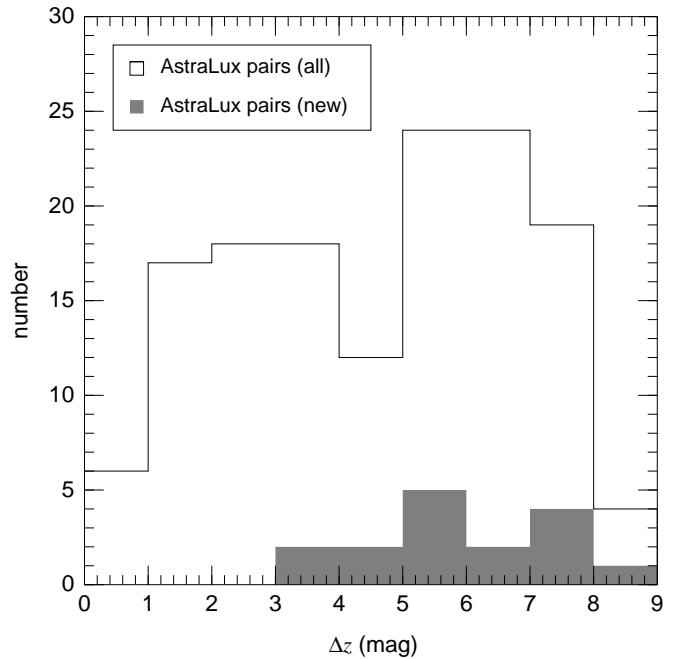}
\caption{\deltam\ histogram for all and for new pairs detected with AstraLux. 
As in the two previous figures,
pairs are selected with one of the components being always the brightest star in the field.
}
\label{dm}
\end{figure}

\begin{enumerate}
  \item The most important one is the contribution from short-period spectroscopic binaries (SBs), many of which have 
        separations well beyond AstraLux capabilities. \citet{Masoetal98} already found multiplicity fractions of
	30-50\% among SBs and recent works have found many new discoveries. Indeed, within GOSSS we have found several
	tens of new SBs (Barb{\'a} et al., in preparation), so the SB fraction should be even higher.
  \item $\deltam \sim 8-9$ (the limit here) is not the expected limit for bound companions 
        \citep{Turnetal08,Maizetal10a}. Also, a look at Fig.~\ref{dm} shows that new discoveries tend to be produced at
	large \deltam, indicating that widening the search in \deltam\ will also likely increase the multiplicity
	fraction.
  \item As previously mentioned, our numbers for separations in the 0\farcs1-2\arcsec\ range must depend on completeness
        corrections. Also note in Fig.~\ref{sep} that several new AstraLux pairs are found at low separations.
  \item Finally, there is the issue of the outer limit for bound pairs. 10\arcsec\ at the typical distances of 1-2 kpc 
        for the objects in our sample is 10\,000-20\,000 AU, which is close to the limit one would expect for bound 
	pairs involving a massive star\footnote{Not only for dynamical reasons but also because for larger separations 
        the system would have time for only one or a few orbits before the central star explodes, hence questioning 
	whether we should consider such a system a binary or not.} Therefore, we can consider that the completeness 
	correction due to possible large-separation missed systems is likely small. 
\end{enumerate}

	With the above caveats in mind I derive a number for the multiplicity fraction for objects with separations of 
0\farcs1-14\arcsec\ with $\deltaz < 8$. The rough percentage (66 fields with two or more components out of a total of 138) 
yields $48\pm 4\%$. That value needs to be corrected for two reasons: First, we should consider that $\approx 22\%$ 
($\sim 20$ cases) of the large separation pairs are not bound\footnote{Some of those pairs will be in systems with two 
or more pairs so the
reduction is not as drastic as if they were all in single-pair systems.}. Second, we have estimated that there
are $\sim 30$ undetected pairs in the 0\farcs1-2\arcsec\ range. Since those two numbers are similar and they work in
opposite directions, I conclude that approximately 50\% of the objects in the sample are multiple within the range
of separations and \deltaz\ above, with a more precise number depending on further analysis. Of course, the total
multiplicity fraction for massive stars should be much higher.

\begin{acknowledgements}
I would like to thank Felix Hormuth for helping with the use of the AstraLux instrument and pipeline; Brian D. Mason
for providing corrections for some of the data in Table 2; Alfredo Sota for obtaining complementary information with 
long-slit optical spectroscopy; 
the IAA Director, Paco T.D.W. Prada, for granting the time for this project;
and Nolan R. Walborn and the referee, Doug Gies, for useful comments that improved the manuscript.
This research was conducted using observing time provided by the IAA-CSIC Director Guaranteed Time for the Calar Alto 
Observatory. Support for this work was provided by [a] the Spanish Government Ministerio de Ciencia e Innovaci\'on through 
grants AYA2007-64052, the Ram\'on y Cajal Fellowship program, and FEDER funds; [b] the Junta de Andaluc\'{\i}a
grant P08-TIC-4075; and [c] NASA through grant GO-10602 from the Space Telescope Science Institute, which is operated by
the Association of Universities for Research in Astronomy Inc., under NASA contract NAS~5-26555. This research has made 
use of [a] Aladin \citep{Bonnetal00}; [b] the SIMBAD database, operated 
at CDS, Strasbourg, France; [c] data products from the Two Micron All Sky Survey, which is a joint project of the 
University of Massachusetts and the Infrared Processing and Analysis Center/California Institute of Technology, 
funded by the National Aeronautics and Space Administration and the National Science Foundation; and [d] the 
Tycho-2 catalog \citep{Hogetal00a}.
\end{acknowledgements}

\bibliographystyle{aa}
\bibliography{general}

\addtocounter{table}{1}
\longtabL{3}{
\begin{landscape}
\begin{longtable}{llccr@{$\pm$}lr@{$\pm$}lcrr@{$\pm$}lrr}
\caption{AstraLux measurements of systems with multiple components.}
\label{pair_data} \\
\hline
\hline
\multicolumn{1}{c}{System name} & \multicolumn{1}{c}{WDS} & \multicolumn{1}{c}{Pair} & MJD & \multicolumn{2}{c}{Separation} & \multicolumn{2}{c}{Orientation} & pair    & \multicolumn{1}{c}{$\Delta V_{\rm Ty}$} & \multicolumn{2}{c}{$\Delta z$} & \multicolumn{1}{c}{$\Delta H_{\rm 2M}$} & \multicolumn{1}{c}{$\Delta K_{\rm 2M}$} \\
                                &                         &                          &     & \multicolumn{2}{c}{(\arcsec)}  & \multicolumn{2}{c}{(degrees)}   & in WDS? & \multicolumn{1}{c}{(mag.)}              & \multicolumn{2}{c}{(mag.)}     & \multicolumn{1}{c}{(mag.)}              & \multicolumn{1}{c}{(mag.)}              \\
\multicolumn{14}{c}{} \\
\hline
\multicolumn{14}{c}{} \\
\endfirsthead
\multicolumn{14}{c}{{\bfseries \tablename\ \thetable{}}} \\
\multicolumn{14}{c}{(Continued)} \\
\multicolumn{14}{c}{} \\
\hline
\hline
\multicolumn{1}{c}{System name} & \multicolumn{1}{c}{WDS} & \multicolumn{1}{c}{Pair} & MJD & \multicolumn{2}{c}{Separation} & \multicolumn{2}{c}{Orientation} & pair    & \multicolumn{1}{c}{$\Delta V_{\rm Ty}$} & \multicolumn{2}{c}{$\Delta z$} & \multicolumn{1}{c}{$\Delta H_{\rm 2M}$} & \multicolumn{1}{c}{$\Delta K_{\rm 2M}$} \\
                                &                         &                          &     & \multicolumn{2}{c}{(\arcsec)}  & \multicolumn{2}{c}{(degrees)}   & in WDS? & \multicolumn{1}{c}{(mag.)}              & \multicolumn{2}{c}{(mag.)}     & \multicolumn{1}{c}{(mag.)}              & \multicolumn{1}{c}{(mag.)}              \\
\multicolumn{14}{c}{} \\
\hline
\multicolumn{14}{c}{} \\
\endhead
\multicolumn{14}{c}{} \\
\hline
\multicolumn{14}{c}{} \\
\multicolumn{14}{c}{Continued on next page}
\endfoot
\hline
\multicolumn{14}{c}{} \\
\multicolumn{14}{l}{$^a$ Large ($>0.2$ mag) photometric uncertainty. $^b$ Low photometric quality (X flag).}
\endlastfoot
HD 225\,146          & 00040$+$6106 & A-B   & 54483.76 & $ 8.802$ & $ 0.016$ & $ 75.16$ & $ 0.09$ & n & \ldots    & $ 7.64$ & $ 0.06$ &   $ 5.70$ &   $ 5.63$ \\
HD 108               & 00061$+$6341 & A-C   & 54482.78 & $13.281$ & $ 0.018$ & $227.69$ & $ 0.07$ & n & \ldots    & $ 6.76$ & $ 0.07$ &   $ 6.27$ &   $ 6.22$ \\
HD 1337              & 00177$+$5126 & A-C   & 54482.76 & $12.933$ & $ 0.019$ & $103.68$ & $ 0.08$ & n & \ldots    & $ 7.66$ & $ 0.05$ &   $ 6.14$ &   $ 6.12$ \\
HD 5005              & 00528$+$5638 & A-B   & 54416.94 & $ 1.529$ & $ 0.016$ & $ 81.62$ & $ 0.48$ & y &   $ 0.75$ & $ 1.56$ & $ 0.03$ & \ldots    & \ldots    \\
                     &              & A-C   & 54416.94 & $ 3.889$ & $ 0.017$ & $133.12$ & $ 0.42$ & y &   $ 0.31$ & $ 0.63$ & $ 0.02$ &   $ 0.13$ &   $ 0.26$ \\
                     &              & A-D   & 54416.94 & $ 8.902$ & $ 0.022$ & $193.53$ & $ 0.40$ & y &   $ 1.07$ & $ 1.33$ & $ 0.02$ &   $ 1.74$ &   $ 1.77$ \\
HD 8768              & 01281$+$6317 & A-B   & 54761.13 & $ 0.656$ & $ 0.016$ & $ 27.54$ & $ 0.38$ & y & \ldots    & $ 2.55$ & $ 0.02$ & \ldots    & \ldots    \\
                     &              & A-C   & 54761.13 & $ 5.918$ & $ 0.016$ & $ 24.28$ & $ 0.13$ & n & \ldots    & $ 6.47$ & $ 0.03$ & $^a 2.69$ &   $ 5.66$ \\
HD 10\,125           & 01409$+$6410 & A-B   & 54482.82 & $ 0.715$ & $ 0.016$ & $232.52$ & $ 0.38$ & y & \ldots    & $ 3.08$ & $ 0.04$ & \ldots    & \ldots    \\
HD 13\,022           & 02095$+$5847 & A-B   & 54483.78 & $13.382$ & $ 0.016$ & $129.88$ & $ 0.07$ & n & \ldots    & $ 5.56$ & $ 0.14$ &   $ 4.08$ &   $ 3.93$ \\
HD 13\,268           & 02115$+$5610 & A-B   & 54761.14 & $10.734$ & $ 0.017$ & $200.78$ & $ 0.08$ & n & \ldots    & $ 7.85$ & $ 0.06$ &   $ 7.06$ &   $ 6.95$ \\
BD +60 499           & 02323$+$6133 & A-B   & 54483.82 & $ 4.617$ & $ 0.016$ & $226.30$ & $ 0.15$ & n & \ldots    & $ 4.83$ & $ 0.02$ & $^a 0.90$ & $^a 0.89$ \\
                     &              & A-C   & 54483.82 & $14.054$ & $ 0.017$ & $119.60$ & $ 0.06$ & n & \ldots    & $ 5.91$ & $ 0.14$ &   $ 4.39$ &   $ 4.30$ \\
BD +60 501           & 02326$+$6128 & A-B   & 54483.83 & $12.024$ & $ 0.016$ & $ 57.19$ & $ 0.07$ & n & \ldots    & $ 6.71$ & $ 0.05$ &   $ 5.62$ &   $ 5.45$ \\
                     &              & A-C   & 54483.83 & $ 7.517$ & $ 0.017$ & $296.05$ & $ 0.11$ & n & \ldots    & $ 7.53$ & $ 0.06$ &   $ 5.36$ &   $ 5.28$ \\
HD 15\,558           & 02327$+$6127 & A-B   & 54483.84 & $ 9.883$ & $ 0.016$ & $ 96.79$ & $ 0.08$ & y &   $ 2.10$ & $ 2.81$ & $ 0.05$ &   $ 2.77$ &   $ 2.76$ \\
                     &              & A-F   & 54483.84 & $ 4.536$ & $ 0.016$ & $263.27$ & $ 0.15$ & y & \ldots    & $ 6.29$ & $ 0.09$ & \ldots    & \ldots    \\
                     &              & A-I   & 54483.84 & $10.626$ & $ 0.016$ & $ 53.15$ & $ 0.08$ & n & \ldots    & $ 7.67$ & $ 0.05$ & $^a 5.70$ &   $ 6.08$ \\
HD 16\,429           & 02407$+$6117 & Aa-Ab & 54483.87 & $ 0.295$ & $ 0.016$ & $ 91.73$ & $ 0.52$ & y & \ldots    & $ 2.82$ & $ 0.61$ & \ldots    & \ldots    \\
                     &              & Aa-B  & 54483.87 & $ 6.777$ & $ 0.016$ & $190.28$ & $ 0.11$ & y &   $ 1.97$ & $ 2.49$ & $ 0.22$ &   $ 2.37$ &   $ 2.44$ \\
                     &              & Aa-D  & 54483.87 & $ 2.947$ & $ 0.104$ & $112.54$ & $ 2.33$ & y & \ldots    & $ 6.66$ & $ 1.33$ & \ldots    & \ldots    \\
HD 16\,832           & 02442$+$5639 & A-B   & 54483.81 & $12.085$ & $ 0.016$ & $353.63$ & $ 0.07$ & n & \ldots    & $ 7.06$ & $ 0.04$ &   $ 5.74$ &   $ 5.55$ \\
                     &              & A-C   & 54483.81 & $ 4.761$ & $ 0.017$ & $335.52$ & $ 0.15$ & n & \ldots    & $ 8.10$ & $ 0.08$ & \ldots    & \ldots    \\
HD 17\,505           & 02511$+$6025 & A-B   & 54483.90 & $ 2.153$ & $ 0.016$ & $ 92.67$ & $ 0.24$ & y &   $ 1.70$ & $ 1.75$ & $ 0.03$ & \ldots    & \ldots    \\
                     &              & A-I   & 54483.90 & $ 4.676$ & $ 0.016$ & $144.09$ & $ 0.15$ & y & \ldots    & $ 6.64$ & $ 0.02$ & \ldots    & \ldots    \\
                     &              & A-J   & 54483.90 & $ 8.912$ & $ 0.016$ & $ 96.13$ & $ 0.09$ & n & \ldots    & $ 6.52$ & $ 0.02$ & $^a 4.88$ & $^a 5.73$ \\
                     &              & A-K   & 54483.90 & $13.698$ & $ 0.017$ & $313.82$ & $ 0.06$ & n & \ldots    & $ 7.00$ & $ 0.06$ &   $ 6.17$ &   $ 6.00$ \\
                     &              & A-L   & 54483.90 & $13.531$ & $ 0.017$ & $224.52$ & $ 0.07$ & n & \ldots    & $ 7.76$ & $ 0.05$ &   $ 6.72$ &   $ 6.54$ \\
HD 17\,520           & 02512$+$6023 & A-B   & 54483.88 & $ 0.316$ & $ 0.016$ & $297.38$ & $ 0.61$ & y & \ldots    & $ 0.67$ & $ 0.07$ & \ldots    & \ldots    \\
                     &              & A-C   & 54483.88 & $10.702$ & $ 0.016$ & $ 20.42$ & $ 0.08$ & y & \ldots    & $ 2.81$ & $ 0.05$ &   $ 3.15$ &   $ 3.24$ \\
                     &              & A-O   & 54483.88 & $ 5.122$ & $ 0.016$ & $299.41$ & $ 0.14$ & n & \ldots    & $ 5.76$ & $ 0.06$ & \ldots    & \ldots    \\
BD +60 586           & 02542$+$6039 & A-B   & 54483.82 & $ 7.131$ & $ 0.016$ & $233.50$ & $ 0.11$ & y & \ldots    & $ 4.03$ & $ 0.14$ &   $ 3.58$ &   $ 3.48$ \\
                     &              & A-C   & 54483.82 & $15.946$ & $ 0.017$ & $ 52.72$ & $ 0.06$ & y & \ldots    & $ 3.41$ & $ 0.13$ &   $ 3.12$ &   $ 3.07$ \\
                     &              & C-F   & 54483.82 & $ 1.811$ & $ 0.018$ & $171.95$ & $ 0.37$ & n & \ldots    & $ 4.55$ & $ 0.20$ & \ldots    & \ldots    \\
                     &              & A-D   & 54483.82 & $ 6.956$ & $ 0.016$ & $105.37$ & $ 0.11$ & n & \ldots    & $ 7.11$ & $ 0.05$ &   $ 4.91$ & $^a 4.80$ \\
                     &              & A-E   & 54483.82 & $19.308$ & $ 0.017$ & $ 63.53$ & $ 0.05$ & n & \ldots    & $ 7.10$ & $ 0.08$ &   $ 5.23$ &   $ 4.96$ \\
                     &              & A-G   & 54483.82 & $ 9.284$ & $ 0.017$ & $ 20.48$ & $ 0.09$ & n & \ldots    & $ 8.57$ & $ 0.09$ & $^a 5.26$ & $^a 5.21$ \\
                     &              & A-H   & 54483.82 & $ 9.984$ & $ 0.017$ & $343.09$ & $ 0.09$ & n & \ldots    & $ 8.73$ & $ 0.08$ & $^a 6.24$ & $^a 6.28$ \\
HD 18\,326           & 02594$+$6034 & A-B   & 54483.89 & $ 2.451$ & $ 0.144$ & $355.97$ & $ 3.14$ & y & \ldots    & $ 8.15$ & $ 1.34$ & \ldots    & \ldots    \\
                     &              & A-C   & 54483.89 & $ 8.039$ & $ 0.016$ & $353.98$ & $ 0.10$ & n & \ldots    & $ 5.00$ & $ 0.04$ &   $ 4.54$ &   $ 4.59$ \\
                     &              & A-D   & 54483.89 & $12.443$ & $ 0.016$ & $327.28$ & $ 0.07$ & n & \ldots    & $ 5.72$ & $ 0.05$ &   $ 5.08$ &   $ 5.19$ \\
                     &              & A-E   & 54483.89 & $ 8.371$ & $ 0.016$ & $248.60$ & $ 0.09$ & n & \ldots    & $ 6.03$ & $ 0.09$ &   $ 5.02$ &   $ 5.00$ \\
                     &              & E-F   & 54483.89 & $ 0.992$ & $ 0.016$ & $ 24.54$ & $ 0.35$ & n & \ldots    & $ 0.57$ & $ 0.12$ & \ldots    & \ldots    \\
HD 24\,431           & 03556$+$5238 & A-B   & 54483.91 & $ 0.720$ & $ 0.016$ & $177.00$ & $ 0.37$ & y & \ldots    & $ 2.80$ & $ 0.02$ & \ldots    & \ldots    \\
NSV 1458 - SZ Cam    & 04078$+$6220 & A-B   & 54760.15 & $ 6.835$ & $ 0.016$ & $256.38$ & $ 0.11$ & y & \ldots    & $ 5.59$ & $ 0.02$ &   $ 3.98$ & $^a 3.74$ \\
                     &              & A-C   & 54760.15 & $11.173$ & $ 0.017$ & $  0.02$ & $ 0.08$ & y & \ldots    & $ 4.56$ & $ 0.02$ &   $ 4.08$ &   $ 4.03$ \\
                     &              & A-D   & 54760.15 & $14.450$ & $ 0.018$ & $131.92$ & $ 0.07$ & y & \ldots    & $ 6.29$ & $ 0.02$ &   $ 5.55$ &   $ 5.50$ \\
                     &              & Ea-Eb & 54483.91 & $ 0.103$ & $ 0.016$ & $290.53$ & $ 1.86$ & y & \ldots    & $ 0.02$ & $ 0.08$ & \ldots    & \ldots    \\
                     &              & Ea-A  & 54483.91 & $17.840$ & $ 0.017$ & $124.79$ & $ 0.05$ & y &   $-0.03$ & $-1.03$ & $ 0.54$ &   $-0.15$ &   $-0.13$ \\
                     &              & Ea-B  & 54483.91 & $14.241$ & $ 0.017$ & $145.84$ & $ 0.06$ & y & \ldots    & $ 4.86$ & $ 0.31$ &   $ 3.83$ & $^a 3.60$ \\
                     &              & Ea-C  & 54483.91 & $14.674$ & $ 0.017$ & $ 86.21$ & $ 0.06$ & y & \ldots    & $ 4.03$ & $ 0.30$ &   $ 3.93$ &   $ 3.90$ \\
HD 34\,656           & 05207$+$3726 & Aa-Ab & 54483.92 & $ 0.373$ & $ 0.017$ & $276.24$ & $ 0.94$ & y & \ldots    & $ 3.13$ & $ 0.07$ & \ldots    & \ldots    \\
                     &              & Aa-B  & 54483.92 & $ 1.948$ & $ 0.018$ & $ 48.49$ & $ 0.36$ & y & \ldots    & $ 7.51$ & $ 0.19$ & \ldots    & \ldots    \\
                     &              & Aa-E  & 54483.92 & $ 8.683$ & $ 0.016$ & $ 12.83$ & $ 0.09$ & y & \ldots    & $ 6.77$ & $ 0.04$ & \ldots    & \ldots    \\
HDE 242\,926         & 05227$+$3319 & A-B   & 54483.99 & $ 1.693$ & $ 0.019$ & $171.86$ & $ 0.33$ & n & \ldots    & $ 4.89$ & $ 0.23$ & \ldots    & \ldots    \\
HDE 242\,935         & 05228$+$3325 & A-B   & 54483.99 & $ 1.081$ & $ 0.016$ & $194.60$ & $ 0.32$ & y &   $ 0.77$ & $ 0.80$ & $ 0.01$ & \ldots    & \ldots    \\
                     &              & A-C   & 54483.99 & $ 9.736$ & $ 0.016$ & $114.75$ & $ 0.08$ & y & \ldots    & $ 3.30$ & $ 0.01$ &   $ 3.38$ &   $ 3.39$ \\
                     &              & A-D   & 54483.99 & $10.512$ & $ 0.016$ & $332.65$ & $ 0.08$ & y & \ldots    & $ 2.53$ & $ 0.01$ &   $ 2.55$ &   $ 2.51$ \\
HD 35\,619           & 05276$+$3445 & A-B   & 54484.00 & $ 2.772$ & $ 0.016$ & $307.29$ & $ 0.20$ & y & \ldots    & $ 2.88$ & $ 0.01$ & \ldots    & \ldots    \\
LY Aur               & 05297$+$3523 & A-B   & 54483.92 & $ 0.598$ & $ 0.016$ & $253.69$ & $ 0.39$ & y & \ldots    & $ 1.87$ & $ 0.02$ & \ldots    & \ldots    \\
$\delta$ Ori         & 05320$-$0018 & Aa-Ab & 54482.84 & $ 0.325$ & $ 0.016$ & $132.66$ & $ 0.47$ & y &   $ 1.35$ & $ 1.48$ & $ 0.02$ & \ldots    & \ldots    \\
HD 36\,483           & 05337$+$3628 & A-B   & 54484.01 & $ 9.900$ & $ 0.016$ & $265.53$ & $ 0.08$ & y & \ldots    & $ 4.81$ & $ 0.01$ &   $ 4.24$ &   $ 4.18$ \\
$\lambda$ Ori        & 05351$+$0956 & A-B   & 54482.90 & $ 4.342$ & $ 0.016$ & $ 43.80$ & $ 0.16$ & y &   $ 1.94$ & $ 1.91$ & $ 0.02$ & \ldots    & \ldots    \\
$\theta^1$ Ori       & 05353$-$0523 & C-A   & 54487.04 & $12.858$ & $ 0.019$ & $311.84$ & $ 0.08$ & y &   $ 1.49$ & $ 1.64$ & $ 0.05$ &   $ 1.17$ &   $ 1.27$ \\
                     &              & C-Ba  & 54487.04 & $16.855$ & $ 0.021$ & $342.80$ & $ 0.07$ & y & $^a 2.43$ & $ 3.02$ & $ 0.04$ &   $ 1.67$ &   $ 1.60$ \\
                     &              & Ba-Bb & 54487.04 & $ 0.996$ & $ 0.016$ & $252.63$ & $ 0.36$ & y & \ldots    & $ 3.02$ & $ 0.02$ & \ldots    & \ldots    \\
                     &              & C-D   & 54487.04 & $13.401$ & $ 0.019$ & $ 61.86$ & $ 0.08$ & y &   $ 1.32$ & $ 1.49$ & $ 0.07$ &   $ 1.26$ &   $ 1.35$ \\
                     &              & C-E   & 54487.04 & $16.630$ & $ 0.021$ & $321.42$ & $ 0.07$ & y & \ldots    & $ 3.63$ & $ 0.05$ &   $ 1.52$ &   $ 1.58$ \\
                     &              & C-H   & 54487.04 & $ 9.426$ & $ 0.018$ & $272.47$ & $ 0.10$ & y & \ldots    & $ 6.81$ & $ 0.12$ &   $ 3.93$ &   $ 3.49$ \\
                     &              & H-I   & 54487.04 & $ 1.547$ & $ 0.017$ & $270.16$ & $ 0.35$ & y & \ldots    & $ 0.34$ & $ 0.21$ & \ldots    & \ldots    \\
                     &              & C-F   & 54487.04 & $ 4.555$ & $ 0.016$ & $120.70$ & $ 0.16$ & y & \ldots    & $ 4.72$ & $ 0.17$ & \ldots    & \ldots    \\
                     &              & C-G   & 54487.04 & $ 7.731$ & $ 0.018$ & $ 33.28$ & $ 0.12$ & y & \ldots    & $ 7.80$ & $ 0.05$ & $^a 4.26$ &   $ 4.05$ \\
                     &              & C-Z   & 54487.04 & $ 6.879$ & $ 0.018$ & $337.04$ & $ 0.13$ & n & \ldots    & $ 7.66$ & $ 0.04$ & \ldots    & \ldots    \\
$\theta^2$ Ori       & 05354$-$0525 & Aa-Ab & 54482.92 & $ 0.396$ & $ 0.016$ & $295.12$ & $ 0.56$ & y & \ldots    & $ 2.62$ & $ 0.04$ & \ldots    & \ldots    \\
$\iota$ Ori          & 05354$-$0555 & A-B   & 54482.94 & $11.319$ & $ 0.018$ & $141.14$ & $ 0.08$ & y & \ldots    & $ 4.51$ & $ 0.01$ & $^a 3.96$ & $^a 3.95$ \\
HD 36\,879           & 05357$+$2124 & A-B   & 54484.02 & $ 9.387$ & $ 0.016$ & $267.46$ & $ 0.09$ & n & \ldots    & $ 3.31$ & $ 0.01$ &   $ 1.47$ &   $ 1.20$ \\
$\sigma$ Ori         & 05387$-$0236 & A-B   & 54482.83 & $ 0.260$ & $ 0.016$ & $ 94.47$ & $ 0.59$ & y & \ldots    & $ 1.57$ & $ 0.30$ & \ldots    & \ldots    \\
                     &              & A-C   & 54482.83 & $11.443$ & $ 0.018$ & $237.89$ & $ 0.08$ & y & \ldots    & $ 5.23$ & $ 0.29$ & $^a 4.47$ &   $ 4.64$ \\
                     &              & A-D   & 54482.83 & $13.031$ & $ 0.019$ & $ 83.99$ & $ 0.08$ & y &   $ 2.79$ & $ 3.05$ & $ 0.28$ & $^a 2.58$ &   $ 2.77$ \\
                     &              & A-G   & 54482.83 & $ 3.224$ & $ 0.045$ & $ 20.34$ & $ 0.65$ & y & \ldots    & $ 7.21$ & $ 0.34$ & \ldots    & \ldots    \\
HD 37\,366           & 05394$+$3053 & A-B   & 54483.07 & $ 0.593$ & $ 0.016$ & $295.47$ & $ 0.50$ & y & \ldots    & $ 3.73$ & $ 0.15$ & \ldots    & \ldots    \\
                     &              & A-C   & 54483.07 & $11.856$ & $ 0.018$ & $ 44.29$ & $ 0.08$ & n & \ldots    & $ 4.29$ & $ 0.02$ &   $ 3.79$ &   $ 3.67$ \\
$\zeta$ Ori          & 05407$-$0157 & A-B   & 54482.85 & $ 2.424$ & $ 0.016$ & $165.70$ & $ 0.22$ & y &   $ 1.81$ & $ 2.26$ & $ 0.02$ & \ldots    & \ldots    \\
HD 41\,161           & 06059$+$4815 & A-B   & 54484.08 & $10.277$ & $ 0.016$ & $269.14$ & $ 0.08$ & y & \ldots    & $ 5.41$ & $ 0.02$ &   $ 4.32$ &   $ 4.25$ \\
HDE 254\,755         & 06185$+$2241 & A-B   & 54761.14 & $ 0.188$ & $ 0.016$ & $ 53.38$ & $ 0.62$ & y & \ldots    & $ 2.09$ & $ 0.05$ & \ldots    & \ldots    \\
HD 44\,811           & 06246$+$1942 & A-B   & 54484.04 & $ 6.211$ & $ 0.016$ & $139.95$ & $ 0.12$ & y & $^a 4.03$ & $ 3.71$ & $ 0.01$ &   $ 3.38$ &   $ 3.23$ \\
HD 46\,056           & 06313$+$0450 & A-B   & 54487.06 & $10.419$ & $ 0.018$ & $336.50$ & $ 0.09$ & y & \ldots    & $ 2.78$ & $ 0.10$ &   $ 2.48$ &   $ 2.46$ \\
HD 46\,150           & 06319$+$0457 & A-B   & 54487.07 & $ 3.516$ & $ 0.017$ & $286.53$ & $ 0.20$ & y & \ldots    & $ 4.33$ & $ 0.04$ & \ldots    & \ldots    \\
                     &              & A-C   & 54487.07 & $ 6.853$ & $ 0.017$ & $319.67$ & $ 0.12$ & y & \ldots    & $ 5.98$ & $ 0.10$ & \ldots    & \ldots    \\
                     &              & A-D   & 54487.07 & $12.519$ & $ 0.019$ & $289.42$ & $ 0.08$ & y & \ldots    & $ 5.36$ & $ 0.02$ &   $ 4.53$ &   $ 4.43$ \\
                     &              & A-L   & 54487.07 & $10.404$ & $ 0.018$ & $ 46.79$ & $ 0.09$ & n & \ldots    & $ 6.29$ & $ 0.11$ &   $ 4.78$ &   $ 4.63$ \\
HD 46\,202           & 06321$+$0458 & D-E   & 54484.05 & $ 3.707$ & $ 0.016$ & $262.90$ & $ 0.18$ & n & \ldots    & $ 3.11$ & $ 0.02$ & $^a 0.69$ &   $ 0.32$ \\
HD 47\,129           & 06374$+$0608 & A-B   & 54487.09 & $ 1.156$ & $ 0.029$ & $252.48$ & $ 1.11$ & y & \ldots    & $ 6.28$ & $ 0.26$ & \ldots    & \ldots    \\
HD 47\,032           & 06388$+$0442 & A-B   & 54761.15 & $ 3.490$ & $ 0.016$ & $258.84$ & $ 0.19$ & n & \ldots    & $ 5.80$ & $ 0.06$ & \ldots    & \ldots    \\
15 Mon               & 06410$+$0954 & Aa-Ab & 54482.93 & $ 0.128$ & $ 0.016$ & $260.94$ & $ 1.47$ & y & \ldots    & $ 1.43$ & $ 0.13$ & \ldots    & \ldots    \\
                     &              & Aa-B  & 54482.93 & $ 2.976$ & $ 0.017$ & $213.50$ & $ 0.21$ & y &   $ 3.15$ & $ 3.23$ & $ 0.16$ & \ldots    & \ldots    \\
HD 48\,279           & 06427$+$0143 & A-B   & 54760.16 & $ 6.860$ & $ 0.016$ & $193.85$ & $ 0.11$ & y &   $ 2.72$ & $ 2.36$ & $ 0.06$ &   $ 1.62$ &   $ 1.56$ \\
HD 52\,533           & 07015$-$0307 & A-B   & 54761.16 & $ 2.676$ & $ 0.019$ & $186.90$ & $ 0.28$ & y & \ldots    & $ 4.89$ & $ 0.07$ & \ldots    & \ldots    \\
                     &              & A-E   & 54761.16 & $11.504$ & $ 0.017$ & $305.64$ & $ 0.08$ & n & \ldots    & $ 5.92$ & $ 0.03$ &   $ 4.99$ &   $ 4.83$ \\
                     &              & A-F   & 54761.16 & $11.281$ & $ 0.017$ & $194.74$ & $ 0.08$ & n & \ldots    & $ 5.78$ & $ 0.03$ &   $ 5.58$ &   $ 5.55$ \\
HD 167\,659          & 18170$-$1858 & A-B   & 54963.15 & $17.531$ & $ 0.034$ & $226.71$ & $ 0.39$ & n & \ldots    & $ 6.10$ & $ 0.02$ &   $ 4.81$ &   $ 4.76$ \\
HD 167\,771          & 18175$-$1828 & A-B   & 54632.08 & $ 8.548$ & $ 0.022$ & $168.78$ & $ 0.40$ & y & \ldots    & $ 5.93$ & $ 0.06$ & \ldots    & \ldots    \\
HD 190\,429          & 20035$+$3601 & A-B   & 54631.95 & $ 1.959$ & $ 0.016$ & $173.80$ & $ 0.46$ & y &   $ 0.64$ & $ 0.61$ & $ 0.01$ & \ldots    & \ldots    \\
WR 133               & 20060$+$3547 & A-B   & 54631.96 & $ 6.855$ & $ 0.020$ & $ 63.36$ & $ 0.40$ & y & \ldots    & $ 5.27$ & $ 0.02$ & \ldots    & \ldots    \\
                     &              & A-C   & 54631.96 & $12.394$ & $ 0.027$ & $ 28.96$ & $ 0.39$ & y & \ldots    & $ 3.46$ & $ 0.01$ &   $ 2.40$ &   $ 2.36$ \\
                     &              & A-D   & 54631.96 & $11.286$ & $ 0.025$ & $300.13$ & $ 0.39$ & y &   $ 2.71$ & $ 2.98$ & $ 0.01$ &   $ 3.00$ &   $ 3.09$ \\
HD 193\,322          & 20181$+$4044 & Aa-Ab & 54632.07 & $ 0.055$ & $ 0.016$ & $119.81$ & $ 2.47$ & y & \ldots    & $-0.04$ & $ 0.19$ & \ldots    & \ldots    \\
                     &              & Aa-B  & 54632.07 & $ 2.719$ & $ 0.017$ & $244.97$ & $ 0.44$ & y &   $ 2.28$ & $ 1.52$ & $ 0.10$ & \ldots    & \ldots    \\
WR 140               & 20205$+$4351 & A-B   & 54632.09 & $ 4.815$ & $ 0.018$ & $210.94$ & $ 0.41$ & y & \ldots    & $ 6.56$ & $ 0.04$ & \ldots    & \ldots    \\
Cyg OB2-4            & 20322$+$4127 & A-B   & 54417.86 & $14.607$ & $ 0.030$ & $327.09$ & $ 0.39$ & n & \ldots    & $ 1.96$ & $ 0.01$ &   $ 2.00$ &   $ 1.97$ \\
                     &              & A-C   & 54417.86 & $18.265$ & $ 0.035$ & $348.72$ & $ 0.39$ & n & \ldots    & $ 3.17$ & $ 0.01$ &   $ 3.12$ &   $ 3.08$ \\
Cyg OB2-5            & 20324$+$4118 & A-B   & 54417.78 & $ 0.934$ & $ 0.016$ & $ 54.42$ & $ 0.53$ & y & $^a 2.55$ & $ 3.02$ & $ 0.03$ & \ldots    & \ldots    \\
                     &              & A-C   & 54417.78 & $20.472$ & $ 0.039$ & $323.18$ & $ 0.39$ & n & \ldots    & $ 6.02$ & $ 0.02$ &   $ 6.20$ &   $ 6.33$ \\
                     &              & A-D   & 54417.78 & $ 5.532$ & $ 0.020$ & $225.25$ & $ 0.41$ & n & \ldots    & $ 7.38$ & $ 0.15$ & \ldots    & \ldots    \\
Cyg OB2-22           & 20331$+$4113 & A-Ba  & 54416.78 & $ 1.536$ & $ 0.016$ & $145.88$ & $ 0.48$ & y & \ldots    & $ 0.55$ & $ 0.04$ & \ldots    & \ldots    \\
                     &              & Ba-Bb & 54416.78 & $ 0.257$ & $ 0.020$ & $179.89$ & $ 2.72$ & n & \ldots    & $ 3.13$ & $ 0.50$ & \ldots    & \ldots    \\
                     &              & A-C   & 54416.78 & $20.398$ & $ 0.039$ & $151.85$ & $ 0.39$ & n & \ldots    & $ 1.32$ & $ 0.03$ &   $ 1.60$ &   $ 1.54$ \\
                     &              & C-D   & 54416.78 & $11.179$ & $ 0.025$ & $ 30.92$ & $ 0.39$ & n & \ldots    & $ 0.51$ & $ 0.03$ &   $ 0.77$ &   $ 0.86$ \\
                     &              & C-E   & 54416.78 & $ 2.822$ & $ 0.017$ & $218.05$ & $ 0.44$ & n & \ldots    & $ 1.69$ & $ 0.02$ & \ldots    & \ldots    \\
                     &              & E-F   & 54416.78 & $ 1.864$ & $ 0.017$ & $189.13$ & $ 0.48$ & n & \ldots    & $ 1.77$ & $ 0.10$ & \ldots    & \ldots    \\
                     &              & C-G   & 54416.78 & $ 3.862$ & $ 0.017$ & $ 31.90$ & $ 0.42$ & n & \ldots    & $ 3.93$ & $ 0.05$ & \ldots    & \ldots    \\
                     &              & C-H   & 54416.78 & $ 7.892$ & $ 0.021$ & $101.71$ & $ 0.40$ & n & \ldots    & $ 4.27$ & $ 0.04$ &   $ 4.31$ &   $ 4.34$ \\
                     &              & C-I   & 54416.78 & $10.416$ & $ 0.025$ & $ 58.46$ & $ 0.39$ & n & \ldots    & $ 4.56$ & $ 0.06$ & \ldots    & \ldots    \\
Cyg OB2-9            & 20332$+$4115 & A-B   & 54417.82 & $21.005$ & $ 0.040$ & $106.35$ & $ 0.39$ & n & \ldots    & $ 6.41$ & $ 0.03$ &   $ 5.31$ &   $ 4.86$ \\
Cyg OB2-7            & 20332$+$4120 & A-B   & 54417.87 & $18.623$ & $ 0.036$ & $104.23$ & $ 0.39$ & n & \ldots    & $ 2.05$ & $ 0.01$ &   $ 2.12$ &   $ 2.11$ \\
                     &              & A-C   & 54417.87 & $15.676$ & $ 0.032$ & $106.03$ & $ 0.39$ & n & \ldots    & $ 5.87$ & $ 0.03$ & \ldots    & \ldots    \\
                     &              & A-D   & 54417.87 & $20.400$ & $ 0.039$ & $ 79.06$ & $ 0.39$ & n & \ldots    & $ 5.93$ & $ 0.06$ &   $ 5.18$ & $^a 5.11$ \\
Cyg OB2-8            & 20333$+$4119 & A-B   & 54417.83 & $ 9.441$ & $ 0.023$ & $202.11$ & $ 0.39$ & y &   $ 1.57$ & $ 1.29$ & $ 0.01$ &   $ 1.04$ &   $ 1.07$ \\
                     &              & A-D   & 54417.83 & $18.182$ & $ 0.035$ & $ 50.77$ & $ 0.39$ & y & \ldots    & $ 2.96$ & $ 0.01$ &   $ 2.70$ &   $ 2.74$ \\
                     &              & C-E   & 54417.84 & $10.353$ & $ 0.024$ & $184.45$ & $ 0.39$ & n & \ldots    & $ 5.27$ & $ 0.02$ & $^a 4.82$ &   $ 4.91$ \\
LS III +46 11        & 20352$+$4651 & A-B   & 54417.91 & $22.563$ & $ 0.042$ & $139.77$ & $ 0.39$ & n & \ldots    & $ 5.28$ & $ 0.02$ &   $ 4.86$ & $^b$---   \\
                     &              & A-C   & 54417.91 & $14.259$ & $ 0.030$ & $114.21$ & $ 0.39$ & n & \ldots    & $ 6.13$ & $ 0.04$ &   $ 5.65$ & $^b$---   \\
HD 206\,267          & 21390$+$5729 & A-C   & 54417.93 & $11.824$ & $ 0.026$ & $119.31$ & $ 0.39$ & y &   $ 1.75$ & $ 2.52$ & $ 0.01$ &   $ 2.38$ &   $ 2.38$ \\
HD 210\,809          & 22116$+$5226 & A-B   & 54761.06 & $ 5.858$ & $ 0.017$ & $333.41$ & $ 0.13$ & n & \ldots    & $ 6.06$ & $ 0.04$ &   $ 5.20$ &   $ 5.21$ \\
WR 153ab             & 22188$+$5608 & A-B   & 54761.07 & $10.058$ & $ 0.017$ & $ 35.98$ & $ 0.09$ & n & \ldots    & $ 7.11$ & $ 0.05$ & $^a 6.71$ &   $ 6.77$ \\
DH Cep               & 22469$+$5805 & A-B   & 54761.07 & $ 7.307$ & $ 0.017$ & $246.69$ & $ 0.11$ & n & \ldots    & $ 7.15$ & $ 0.06$ & \ldots    & \ldots    \\
HD 217\,086          & 22568$+$6244 & A-B   & 54761.05 & $ 2.865$ & $ 0.016$ & $355.05$ & $ 0.20$ & y &   $ 3.58$ & $ 3.58$ & $ 0.02$ & \ldots    & \ldots    \\
                     &              & A-C   & 54761.05 & $ 3.261$ & $ 0.017$ & $166.22$ & $ 0.23$ & y & \ldots    & $ 6.40$ & $ 0.09$ & \ldots    & \ldots    \\
                     &              & A-D   & 54761.05 & $11.395$ & $ 0.017$ & $ 52.71$ & $ 0.08$ & n & \ldots    & $ 7.78$ & $ 0.06$ &   $ 6.48$ & $^a 6.04$ \\
HD 218\,195          & 23052$+$5815 & A-B   & 54761.08 & $ 0.919$ & $ 0.016$ & $ 80.49$ & $ 0.35$ & y &   $ 2.64$ & $ 2.56$ & $ 0.02$ & \ldots    & \ldots    \\
                     &              & A-C   & 54761.08 & $ 3.545$ & $ 0.017$ & $304.60$ & $ 0.21$ & n & \ldots    & $ 5.65$ & $ 0.06$ & \ldots    & \ldots    \\
                     &              & A-D   & 54761.08 & $11.524$ & $ 0.017$ & $154.05$ & $ 0.08$ & n & \ldots    & $ 6.46$ & $ 0.02$ &   $ 5.85$ &   $ 5.80$ \\
\end{longtable}
\end{landscape}

}

\end{document}